\documentclass[11pt,a4,twoside]{article}
\usepackage[dvips,a4paper,left=2.5cm,right=2.5cm,top=2.5cm,bottom=2.5cm,headsep=1em]{geometry}
\usepackage{fancyhdr}
\usepackage{titlesec}
\usepackage[latin1]{inputenc}
\usepackage{amsmath,amsfonts,amssymb,array,amsthm}
\usepackage{rotating}
\usepackage[font={small}]{caption}
\usepackage{natbib}
\usepackage{lmodern,slantsc}
\usepackage{multirow}
\usepackage{chngcntr}
\usepackage{placeins}
\usepackage{caption}
\usepackage{enumitem}
\usepackage[breaklinks,
colorlinks=true, linkcolor=blue, citecolor=black, urlcolor=blue,
pdfborder={0 0 0}, pdfpagelabels]{hyperref}
\def\vx{\mathbf x}

\def\v0{\boldsymbol{0}}

\newlength{\FigureHeight}
\newlength{\FigureHeightHalf}

\numberwithin{equation}{section}
\titleformat{\section}
{\large\bfseries}{\thetitle.}{0.5em}{}
\titleformat{\subsection}
{\normalfont\bfseries}{\thetitle.}{0.5em}{}
\titleformat{\subsubsection}
{\normalfont\itshape}{\normalfont\thetitle.}{0.5em}{}
\begin{document}

\title{\vspace{0.0em} Refuting the claim of conformal invariance\\ for a zero-vorticity characteristic equation in 2D turbulence
\\
{\large\it A review of
\href{https://doi.org/10.1088/1751-8121/abe95b}
{{\large J.$\,$Phys.A$\,\,$\textbf{\emph{54}},$\,$438001$\,$(2021)}}}}

\author{Michael Frewer$\,^1$\thanks{Email address for correspondence:
frewer.science@gmail.com}$\:\,$ \& George Khujadze$\,^2$ \\ \\
\small $^1$ Heidelberg, Germany\\
\small $^2$ Chair of Fluid Mechanics, Universit\"at Siegen, 57068
Siegen, Germany}
\date{{\small\today}}
\clearpage \maketitle \thispagestyle{empty}

\vspace{-2.0em}\begin{abstract}

\noindent
Although the current Reply by Grebenev~{\it et~al.}~(\href{https://doi.org/10.1088/1751-8121/abe95b}{2021{\it a}}) makes their original analysis
in\linebreak[4] Grebenev~{\it et~al.}~\href{https://doi.org/10.1088/1751-8121/aa8c69}{(2017)} more transparent, the actual problem remains.
Their claim to have analytically proven conformal invariance in 2D turbulence for a zero-vorticity characteristic equation is not~true.
We refuted this claim in Frewer~\&~Khujadze~(\href{https://doi.org/10.1088/1751-8121/abe95a}{2021{\it a}},\href{https://doi.org/10.1088/1751-8121/abe49e}{{\it b}}),
which we will briefly summarize here again with respect to the presented Reply.
In particular our proof on the symmetry-breaking property of the integral normalization constraint
is misrepresented and misconstrued, especially in their second Reply (Grebenev {\it et al.}, \href{https://doi.org/10.1088/1751-8121/abe49f}{2021{\it b}}).
Although the\linebreak[4] journal's only selected expert reviewer clearly agreed with our proof in his final conclusion,\footnote{Due to copyright reasons we may not directly cite the journal's final review but only paraphrase it. It clearly concludes that the error is on Grebenev's side and not ours, by providing arguments in the same line of reasoning as ours that their integration and symmetry analysis is inconsistent and thus flawed. It further concludes that all our derivations and analyses are well done. In the reviewer's opinion, our work is helpful for readers to realize the applicability of Lie group analysis in turbulence and that it is an important value for documentation.}\linebreak[4] the~journal nevertheless decided to publish the Replies.

\vspace{0.5em}\noindent{\footnotesize{\bf Keywords:} {\it Statistical Physics, Conformal Invariance, Turbulence, Probability Density Functions, Lie Groups, Symmetry Analysis, Integro-Differential Equations,
Closure Problem}}
\end{abstract}

\section{The issue of the mentioned group classification}\label{Sec1}
We agree that a group classification of the first equation of the LMN hierarchy with respect to the vorticity variable $\omega$ leads to a conformal invariant result for the specification $\omega=0$. This is trivial and not disputed, because the intermediate symmetry result for $\xi^3$ of the defining linear equation (Eq.$\,$[4] in \cite{Frewer21.1}), namely the result
\begin{equation}
\partial_{x^1}\xi^3=\partial_{x^2}\xi^3=0,\; \text{or briefly}\;\; \xi^3_1=\xi^3_2=0,
\label{210210:0044}
\end{equation}
is trivially satisfied as a solution constraint for $\xi^3=\xi^3(\vx,\omega)$ with the ansatz
\begin{equation}
\xi^3\Big\vert_{\omega=0}=f(\vx)\cdot \omega\,\Big\vert_{\omega=0},
\label{210208:1754}
\end{equation}
where $f$ is some arbitrary function in the spatial coordinates $\vx=(x^1,x^2)$ --- sure, when including also the two nonlocal equations (Eqs.$\,$[5-6] in \cite{Frewer21.1}) into the symmetry analysis, then $f$ reduces to the harmonic function
\begin{equation}
f(\vx)=6c^{11}(\vx),
\label{210208:1755}
\end{equation}
as correctly described in \cite{Grebenev17,Grebenev21.1}. We therefore agree with what is said~in the first part of point (a) for Eqs.$\,$[10-11] in \cite{Grebenev21.1}, but not for what is said in~the second part, particularly that

{\it ``Frewer and Khujadze did not consider the above classification problem"},

\noindent or later in point (d) that

{\it ``neither differential nor integral consequences of this [LMN] equation with respect to $\omega$\linebreak[4]\indent\phantom{``}can be considered"}.

\vspace{1em}\noindent
We considered the classification problem,\footnote{Please see the introductory part of Sec.$\,$[2.1] in \cite{Frewer21.1}, that we were fully aware of the group classification result~\eqref{210208:1754} and its consequences in \cite{Grebenev17}.} but we rejected it, because we already knew that the special solution \eqref{210208:1754} stemming from it will only lead to internal inconsistencies when considering the full system:

Firstly, we knew that the system not only consists of Eqs.$\,$[4-6]$\,$\citep{Frewer21.1},\linebreak[4] but that it also includes Eq.$\,$[7]
\begin{equation}
1-\int\! d\omega f_1=0,
\end{equation}
an equation which obviously involves an integration of $\omega$, which again makes it necessary to consider a {\it global} solution of $\xi^3$ that is continuously valid for all $\omega$ (in order to transform $d\omega$ consistently in line with the other Eqs.$\,$[4-6]), and not only a solution that is only valid for the single point $\omega=0$, as the group classification misleadingly suggests. See also the next section for further details.

Secondly, the symmetry analysis itself involves a differential operator $X$ that contains a derivative in $\omega$ (see Eq.$\,$[27] in \cite{Grebenev17})
\begin{equation}
X\,=\, \xi^3\frac{\partial}{\partial\omega}\,+\, Y,
\label{210208:1612}
\end{equation}
where $Y$ denotes all remaining terms of $X$. That is, the search for symmetries and invariant solutions does not involve the function $\xi^3$ alone, but the product combination of $\xi^3$ along with the differential operator $\partial_\omega$ is what matters in the end.

Hence, already these two facts alone obviously make clear that it is absolutely necessary to consider differential and integral consequences with respect to $\omega$, and not to hide them. Only then an overall consistent symmetry analysis can be guaranteed. And it is exactly this approach we follow in \cite{Frewer21.1}. For example, if we want to construct an overall consistent symmetry solution through the differential generator $X$ \eqref{210208:1612} for the special case~$\omega=0$, as intended in \cite{Grebenev17}, then we should consider and evaluate $X\vert_{\omega=0}$~as
\begin{equation}
X\big\vert_{\omega=0}\,= \bigg(\xi^3\cdot\frac{\partial}{\partial\omega}\bigg)\bigg\vert_{\omega=0}\,+\, Y\big\vert_{\omega=0},
\label{210208:1616}
\end{equation}
and not as
\begin{equation}
X\big\vert_{\omega=0}\,=\, \xi^3\Big\vert_{\omega=0}\cdot \frac{\partial}{\partial\omega}\bigg\vert_{\omega=0}+Y\big\vert_{\omega=0}
\;\underset{\eqref{210208:1754}}{=}\; f(\vx)\cdot \omega\,\Big\vert_{\omega=0}
\cdot \frac{\partial}{\partial\omega}\bigg\vert_{\omega=0}+Y\big\vert_{\omega=0}
=\;0+Y\big\vert_{\omega=0},
\label{210208:1643}
\end{equation}
as their group classification misleadingly suggests. The latter result \eqref{210208:1643} is only true and coincides with \eqref{210208:1616} if the derivative $\partial_\omega$ acting on some external function is {\it regular} at $\omega=0$, otherwise the evaluation \eqref{210208:1643} turns into an indeterminate expression where $\xi^3\vert_{\omega=0}\cdot\partial_\omega\vert_{\omega=0}$ does not necessarily evaluate to zero. But here is the problem now: In order to concretely determine this non-zero contribution, one has to evaluate the form \eqref{210208:1616}, but for that one needs a solution for $\xi^3$ that is valid in some {\it region} around $\omega=0$ and not only in the single value $\omega=0$.\linebreak[4]
In other words, for a derivative $\partial_\omega\vert_{\omega=0}$ that evaluates non-regularly on some external function, but gives a regular result when evaluating it as $(\xi^3\partial_\omega)\vert_{\omega=0}$, the discrete $\omega=0$ evaluation of~\eqref{210208:1616} turns into a continuous limit process where thus a solution for $\xi^3$ not only in $\omega=0$, but at least in some infinitesimal region around $\omega=0$ is needed. The group classification result \eqref{210208:1754} is therefore not enough to correctly determine $X\vert_{\omega=0}$ \eqref{210208:1616}, since \eqref{210208:1754} is obviously only valid for the single discrete value $\omega=0$.\footnote{Note that the smoothness axiom of a Lie-group action dictates the solution of $\xi^3$ itself to be continuously differentiable around $\omega=0$. Thus one could drop the discrete evaluation constraint $\omega=0$ in \eqref{210208:1754} to get a smooth function for $\xi^3$ in~$\omega$, but this will break the conformal invariance again, which surely is not in the interest of \cite{Grebenev17}.}

Hence, \eqref{210208:1612} and \eqref{210208:1616} clearly show that differential consequences in general and in particular for $\omega=0$ play an important role and should not be ignored. The same is true for integral consequences as clarified before. They both provide useful information to guarantee an overall consistent symmetry analysis, as explicitly shown in \cite{Frewer21.1},
both for differential and integral consequences with respect to $\omega$, in Secs.$\,$[2.1.1]~\&~[2.4] respectively.

\section{The issue of the normalization constraint (again)}

Our proof in \cite{Frewer21.1}, as well as in \cite{Frewer21.2}, each independently demonstrating the breaking of the conformal group due to the presence of the integral normalization constraint, are clearly misrepresented in \cite{Grebenev21.1,Grebenev21.2}. Their arguments in both Replies on this issue are just repetitions of what has already been said in \cite{Grebenev17}. No new arguments or a counterproof is provided to refute our proofs.
Instead, we see the same arguments again that we have already refuted. For example, their argument given by Eqs.$\,$[5-6] \citep{Grebenev21.1} that the normalization condition
\begin{equation}
\int\! d\omega f_1=1
\label{210209:1315}
\end{equation}
stays invariant under the general space-dependent scaling
\begin{equation}
\omega^*=\mathcal{D}(\vx)\omega,\quad f^*_1=[\mathcal{D}(\vx)]^{-1}f_1,
\label{210209:1635}
\end{equation}
is trivial and not disputed. The problem, however, is that \eqref{210209:1315} is not a stand-alone equation, but is coupled to a larger set of equations (Eqs.$\,$[4-6] in \cite{Frewer21.1})
\begin{equation}
\left.
\begin{aligned}
\frac{\partial f_1}{\partial t}+ \frac{\partial J^1}{\partial x^1}+\frac{\partial J^2}{\partial x^2}=0,&\\[0.5em]
J^1+\frac{1}{2\pi}\int d^2\vx^\prime d\omega^\prime \omega^\prime\frac{x^2-x^{\prime 2}}{|\vx-\vx^\prime|^2}f_2=0,&\\[0.5em]
J^2-\frac{1}{2\pi}\int d^2\vx^\prime d\omega^\prime \omega^\prime\frac{x^1-x^{\prime 1}}{|\vx-\vx^\prime|^2}f_2=0,&
\end{aligned}
~~~\right\}
\label{210209:1614}
\end{equation}
which all need to stay invariant collectively and simultaneously if one is to assign a symmetry to this entire system \eqref{210209:1315} \& \eqref{210209:1614}.

If, for example, one now goes for the admitted group classified invariance \eqref{210208:1754}-\eqref{210208:1755} of the subsystem \eqref{210209:1614}, as chosen in \cite{Grebenev17}, then the corresponding invariance transformation for $\omega$ (in infinitesimal form, up to order $\mathcal{O}(\epsilon^2)$ in the group parameter~$\epsilon$)
\begin{equation}
\omega^*=\omega+\epsilon\cdot\xi^3\Big\vert_{\omega=0} +\mathcal{O}(\epsilon^2),
\label{210209:1641}
\end{equation}
must be made compatible with the invariant transformation rule \eqref{210209:1635} of the subsystem~\eqref{210209:1315}, so that one obtains a valid symmetry transformation that is admitted by the entire
system~\eqref{210209:1315}~\&~\eqref{210209:1614}. But since, according to \eqref{210208:1754}-\eqref{210208:1755}, the above transformation \eqref{210209:1641} is equivalent to
\begin{equation}
\omega^*=\omega+\epsilon\cdot 0 +\mathcal{O}(\epsilon^2),
\end{equation}
and since we are dealing here with a Lie-group transformation, where the first order term in the group parameter $\epsilon$ already determines the full transformation up to all orders, we finally yield
for subsystem \eqref{210209:1614} the following global invariance transformation for $\omega$:
\begin{equation}
\omega^*=\omega.
\label{210209:1734}
\end{equation}
As a result, we see that the invariant transformation \eqref{210209:1635} of subsystem \eqref{210209:1315} in $\omega$ can only be made compatible with the above result \eqref{210209:1734}, if
\begin{equation}
\mathcal{D}(\vx)=1,
\label{210209:1747}
\end{equation}
i.e., if the scaling factor $\mathcal{D}$ is spatially constant. Only then we have a consistent invariant transformation that is admitted by the entire system \eqref{210209:1315} \& \eqref{210209:1614}.
But condition \eqref{210209:1747}, of course, breaks the conformal invariance of this system.\qed

In the second Reply \citep{Grebenev21.2} we face the very same misrepresentation of our proof on the symmetry breaking property of the normalization constraint \eqref{210209:1315}. Particularly in their last paragraph of Sec.$\,$[1], Grebenev {\it et al.} distort our key argument beyond recognition: Claiming that for the case $\omega=0$ in Eq.$\,$[5] \citep{Frewer21.2} the function $6c^{11}(\vx)$ can be replaced by any other function $f(\vx)$, shows that the key aspect of our proof has not been understood. Surprisingly, it even stands in contrast to what they say in their first Reply~\citep{Grebenev21.1}. Because this part in our Eq.$\,$[5] is nothing else than it is based on their group classification result \eqref{210208:1754}-\eqref{210208:1755}.

As stated in our Comment \citep{Frewer21.2}, and as can be clearly seen too, Eq.$\,$[5] is the result for $\xi^3$ exactly as provided by \cite{Grebenev17} for the subsystem~\eqref{210209:1614}:
\begin{equation}
\xi^3=
\begin{cases}
\: 6c^{11}(\vx)\cdot\omega,\;\text{for $\omega=0$,}\\
\: \hspace{1.07cm}c\cdot\omega,\;\text{for $\omega\neq 0$, $c\neq 0$.}
\end{cases}
\label{190714:1937}
\end{equation}
While the branch $\omega=0$ is the admitted group classified invariance \eqref{210208:1754}-\eqref{210208:1755} of sub-\linebreak[4] system~\eqref{210209:1614}, the non-zero branch $\omega\neq 0$ is the admitted standard invariance of \eqref{210209:1614}. The latter invariance is collectively obtained by the full result $\xi^3=6c^{11}(\vx)\omega$ from the last two equations and the full result $\xi^3_1=\xi^3_2=0$ \eqref{210210:0044} from the first equation of \eqref{210209:1614}. Thus, for the case $\omega\neq 0$, its combined result for all three equations of \eqref{210209:1614}, i.e.,
\begin{equation}
\xi^3=6c^{11}(\vx)\omega \quad\text{and}\quad \xi^3_1=\xi^3_2=0,\quad\text{for}\quad\omega\neq 0,
\end{equation}
is therefore
\begin{equation}
\xi^3=c\cdot\omega,
\end{equation}
where $c$ is some arbitrary integration constant, chosen to be non-zero $(c\neq 0)$. This iterated explanation to comprehend \eqref{190714:1937} and the fact that it's simply just the direct consequence of the analysis provided in \cite{Grebenev17}, adds nothing more here than what has already been said in \cite{Frewer21.2}: Result \eqref{190714:1937} is the only unique solution for~$\xi^3$ of the subsystem~\eqref{210209:1614} that is globally valid for all $\omega$. This result is fixed, nothing can be changed or replaced by using other functions as misleadingly claimed by \cite{Grebenev21.2}, since this obviously will only lead to a result for $\xi^3$ that is not a solution anymore to subsystem \eqref{210209:1614}.

The final step is now to realize that since we are operating with Lie-groups, the element~$\xi^3$ must be smooth (continuously differentiable) in $\omega$ in order to not violate the smoothness-axiom of a Lie-group action. Hence, we can conclude from \eqref{190714:1937} that
\begin{equation}
6c^{11}(\vx)=c,
\end{equation}
i.e., the harmonic function $c^{11}$ has to be spatially constant, thus finally breaking again the conformal invariance of the subsystem \eqref{210209:1614}.\qed

As also said in \cite{Frewer21.2}, important to note here is that the above conclusion does not have to involve the normalization constraint \eqref{210209:1315} to demonstrate the breaking of the conformal invariance, and therefore closing the circle to why we rejected the group classification result in the first place, as explained earlier in the first part of Section \ref{Sec1}.

\section{Final remarks}

\subsection{To a minor technical issue} 

In \cite{Grebenev21.1} it is said that one specific formulation of ours inevitably leads to the conclusion that {\it ``the condition $\xi^3\equiv 0$ should be satisfied identically for all $\omega$"} and that we therefore contradict ourselves. However, from the context it is more than clear that the word `all' in our formulation {\it ``that the proposed conformal invariance is not only broken for $\omega=0$, but for all $\omega\in\mathbb{R}$"} has the meaning of `any'. Thus a small irrelevant imprecision in the wording is blown out of proportion here by Grebenev {\it et al.}

\subsection{To the fundamental and crucial issues}

We would like to emphasize one of our remarks in \cite{Frewer21.1} once again. Important to keep in mind is that since we are dealing with \eqref{210209:1315} \& \eqref{210209:1614} with an unclosed statistical system, all invariant transformations examined and discussed here, are not symmetry but only equivalence transformations. Also, any invariant solution that can be obtained with the general Lie-group method may not be a true solution of the LMN dynamics considered here. Without any kind of mathematical or physical modelling of the LMN equations, nothing can be said about its solution structure. In other words, the Lie-group method alone, like any other analytical method, cannot bypass the closure problem of turbulence. For more details, see remark {\it R5} in Sec.$\,$[3] \citep{Frewer21.1} and the references therein.

Also should be clear here that we did not refute or criticize the numerical result of conformal invariance in 2D turbulence as first shown by \cite{Bernard06}, that the large-scale zero-isolines of the vorticity field in the inverse energy cascade are statistically equivalent to Schramm-L\"owner evolution (SLE) curves, i.e., that they can be conformally mapped to a 1D Brownian walk within the numerical and statistical resolution used. What is refuted and criticized herein is only the claim by Grebenev {\it et al.} to have {\it analytically} proven the existence of conformal invariance in 2D turbulence. Conformal invariance in 2D turbulence does
not reveal itself simply by truncating the infinite hierarchy of the defining statistical equations at first order and then by analyzing its resulting unclosed system only by means of a classical Lie-group
symmetry approach, which then, ultimately, even does not determine any symmetries but only a weaker set of equivalence transformations which in the end are also unable to probe the space of physically realisable solutions.

To note is also that no connection to the results of Bernard {\it et al.} can be established, as misleadingly claimed by Grebenev {\it et al.} The results that Grebenev {\it et al.} wanted to produce are unrelated to the results shown by Bernard {\it et al.}: While Grebenev {\it et al.} wanted to prove conformal invariance in unclosed statistical equations detached from any particular solutions, Bernard {\it et al.} shows how the solution of a large-scale zero-vorticity isoline can be conformally mapped to a Brownian walk. The latter is about the growth of a random fractal curve where each incremental step is produced by a conformal transformation characterized by a Brownian motion, and not about conformal equivalences being admitted by unclosed statistical equations as it would have been the result by Grebenev {\it et al.}

%
%
\bibliographystyle{jfm}
\bibliography{Refs}

\end{document}